\documentclass[aps,prl,reprint,superscriptaddress,citeautoscript]{revtex4-2}
\usepackage[version=4]{mhchem}
\usepackage{amsmath, amssymb}  
\usepackage{graphicx}    
\usepackage{lipsum}
\usepackage{color}  
\usepackage[dvipsnames]{xcolor}

\begin{document}


\title{For molecular polaritons, disorder and phonon timescales control the activation of dark states in the thermodynamic limit}
\author{Tianchu Li}
\affiliation{Department of Chemistry, University of Colorado Boulder, Boulder, Colorado 80309, USA\looseness=-1}
\author{Pranay Venkatesh}
\affiliation{Department of Chemistry, University of Colorado Boulder, Boulder, Colorado 80309, USA\looseness=-1}
\author{Qiang Shi}\email{qshi@iccas.ac.cn}
\affiliation{Beijing National Laboratory for
Molecular Sciences, State Key Laboratory for Structural Chemistry of Unstable and Stable Species, Institute of Chemistry, Chinese
Academy of Sciences, Zhongguancun, Beijing 100190, China}
\affiliation{University of Chinese Academy of Sciences,
Beijing 100049, China}

\author{Andr\'es Montoya-Castillo}
\email{Andres.MontoyaCastillo@colorado.edu}
\affiliation{Department of Chemistry, University of Colorado Boulder, Boulder, Colorado 80309, USA\looseness=-1}
\date{\today}

\begin{abstract}
Collective light--matter systems host an extensive manifold of dark states whose role in the emergence of thermodynamic behavior remains poorly understood, especially in the presence of disorder and structured environments. Here, we develop a hybrid matrix product state--hierarchical equations of motion (MPS--HEOM) approach that enables numerically exact simulations of polariton dynamics from a few emitters to the thermodynamic limit under both static and dynamic disorder. This allows us, for the first time, to provide a quantitative and operational answer to the long-standing question of what is the minimum system size required to reach the thermodynamic limit in collective polaritonic systems. By introducing a convergence scale, $N_T$, i.e., the number of molecules required for the photonic dynamics to reach the thermodynamic limit, we show that dynamic disorder generally poses a greater computational challenge than static disorder. We attribute this behavior to the suppression of collective light–matter dynamics by disorder, which dynamically activates non-collective degrees of freedom. We further find that $N_T$ exhibits a turnover behavior as the bath becomes more Markovian, as the bath timescales regulate bright--to--dark energy transfer and the involvement of dark and gray states. Hence, phonon timescales control both the breakdown of collective behavior and the growth of $N_T$. Our results establish the suppression of collective behavior as the key mechanism governing thermodynamic convergence in disordered light–matter systems.
\end{abstract}

\maketitle

\textit{Introduction} --- Strong light-matter coupling can generate polaritons~\cite{hopfield58,agranovich59, snoke17}, quasiparticles that inherit properties from both constituents: the coherence and delocalization of photons, and the nonlinearity and interactions of matter. Polaritons can offer the ability to tune chemical reactivity~\cite{ribeiro18,garcia-vidal21,long2015,ahn2023,babin2023observation,xiang2024,chen2024,koessler2025polariton}, mediate long-range energy transfer~\cite{coles14, xiang2020,zeb22, balasubrahmaniyam23, xu23,de25}, facilitate room temperature lasing~\cite{kena10, plumhof14, cookson17, wei19, keeling20, ojambati24}, and explore nonequilibrium many-body physics in driven dissipative systems via their Bose-Einstein condensation~\cite{rahimi-iman20}. To capitalize on these promises requires predicting polariton properties across realistic settings. However, while current simulation strategies suffice for atomic polaritons~\cite{strashko18, sokolovskii22, holstein40, gelhausen17, kirton17, li25, haken75, kirton18, lindoy24,li2025disorder}, molecular polaritons, where vibrational coupling can provide an orthogonal means to control polaritonic behavior, remain challenging. 

Organic semiconductors are desirable for molecular polariton applications because of their sizable light-matter coupling~\cite{kena10, ribeiro18, hertzog19} and wide range of exciton-vibrational couplings \cite{silinsh94}. Such vibrational coupling can, for example, increase the pumping threshold for polariton lasing~\cite{fowler22}. To capture molecular polariton physics,~\cite{herrera16,feist2015extraordinary} the Holstein--Tavis--Cummings (HTC) Hamiltonian augments the Tavis--Cummings (TC) description~\cite{tavis68} of atomic polaritons, where matter consists of two-level systems (TLSs), with bilinear coupling to phonons that modulate exciton energies. Because vibrational coupling exponentially increases computational complexity, the HTC model poses major challenges for simulation. As such, computational approaches often face a trade-off: methods that capture collective behavior in the thermodynamic limit often neglect dissipation and pumping~\cite{li25,perez23,perez25}, whereas methods that include dissipation and pumping generally sacrifice thermodynamic-limit observables or correlations~\cite{fowler22,arnardottir25,mu25}. Difficulties notwithstanding, these approaches have revealed how vibrational coupling and dissipation reshape the structure and dynamics of molecular polaritons~\cite{du18,ribeiro18a,arnardottir20, li2021cavity, engelhardt2023polariton,lindoy23, catuto2025interplay,liu2025dissecting}.

Today, most polariton experiments and simulations deviate in two major ways: disorder and system size. First, while experiments on molecular polaritons rely on amorphous molecular thin films, most theoretical approaches focus on idealized TC and HTC models free from disorder. Yet, disorder restructures bright and dark states by breaking their original symmetry~\cite{herrera16, feist2015extraordinary}, and is thought to play a crucial role in polariton-modified chemical reactions~\cite{herrera2017dark, garcia2021manipulating, wellnitz2022disorder}, contribute to Rabi-splitting inversion~\cite{cohn2022vibrational, catuto2025interplay}, and tune polariton transport~\cite{hou2020ultralong, xu23, tutunnikov2024characterization}. Second, experiments employ macroscopic samples with generally $>10^5$ molecules whereas most theoretical methods and \textit{ab initio} simulations applicable to disordered cases apply to small systems with $<20$ molecules~\cite{del2018tensor, aroeira2023theoretical, engelhardt2023polariton, tutunnikov2024characterization, li2022qm}. Without knowing how many emitters are needed to reach the thermodynamic limit, it is difficult to say how well conclusions drawn from these works relate to actual experiments, or how to exploit nonlinearities that emerge below this limit \cite{koner2025hidden}. \textit{Hence, what is needed is the ability to interrogate the nonequilibrium evolution of statically and dynamically disordered molecular polaritons below and at the thermodynamic limit.} 

Here, we propose a matrix product state (MPS) based method to achieve this goal. We build our method around the HTC model, integrating the hierarchical equations of motion (HEOM) \cite{tanimura89,tanimura06,tanimura20} into an optimized MPS architecture to treat vibrationally induced dynamic disorder and intermediate system–bath coupling. This approach captures non-Markovian vibrational relaxation alongside Markovian cavity loss and external driving within a unified tensor-network framework. The TC model with static disorder and Lindblad dissipation emerges as a limiting case. For the first time, this framework quantitatively determines how disordered TC and HTC models converge toward the thermodynamic limit, bridging few-emitter and macroscopic regimes. Moreover, we demonstrate how phonon engineering can selectively enhance dissipative pathways that funnel excitations into the dark-state manifold.



\textit{Systems} --- We begin with the TC and HTC Hamiltonians. The TC model describes a set of $N$ identical TLSs (local excitations) coupled to a single photonic mode,
\begin{align}
    H_{\rm TC} = \omega_c a^\dagger a + \sum_{i=1}^N \left[ \frac{\omega_0}{2} \sigma_i^z + \frac{\Omega}{\sqrt{N}} \left( a^\dagger \sigma_i^- + a \sigma_i^+ \right) \right],
\end{align}
where $a^\dagger$ ($a$) denotes the creation (annihilation) operator for the cavity mode, and $\sigma_i^{z,\pm}$ are the Pauli matrices for the $i$th TLS. $\omega_c$ is the cavity frequency, $\omega_0$ is the frequency of the TLSs, and $\Omega$ is the collective light–matter coupling strength (Rabi frequency). The HTC model, 
\begin{equation}
    H_{\rm HTC} = H_{\rm TC} + \sum_{i=1}^N \left( H_B^i + H_I^i \right), 
\end{equation}
includes coupling to vibrational modes, with $H_B^i$ describing a local phonon bath coupled to the $i$th TLS, $H_B^i =\sum_{\alpha} \omega_{i,\alpha}b_\alpha^+b_\alpha$ and the system–bath interaction, $H_I^i = \sum_{\alpha} c_{i,\alpha} \left(b_\alpha^++b_\alpha\right) \otimes \sigma_i^z$, that modulates the local exciton frequency. Here, $b^+_{\alpha}$ ($b_\alpha$) denotes the creation (annihilation) operator for the $\alpha$th vibrational mode coupled to the $i$th TLS, $\omega_{i,\alpha}$ is the mode frequency, and $c_{i,\alpha}$ represents the coupling strength. 

Consistent with previous studies of exciton-phonon coupling in organic molecular polaritons~\cite{mandal2023theoretical}, we consider the continuous vibrational coupling limit characterized by a spectral density, $J_i(\omega) = \pi \sum_{\alpha} c_{i,\alpha}^2 \delta(\omega - \omega_{i,\alpha})$~\cite{weiss12}, which we assume to be identical for all TLSs: $J_i(\omega) = J(\omega)$ for all $i$. We further adopt a Debye-Drude form $J(\omega) = \frac{\eta \gamma \omega}{\omega^2 + \gamma^2}$, where $\eta = 2/\pi \int_0^{\infty} d\omega\ J(\omega)/\omega$ is the reorganization energy that quantifies exciton-vibrational coupling and $\gamma = \tau_B^{-1}$ is the characteristic timescale $\tau_B$ over which the vibrations dissipate energy.

\begin{figure}[htbp]
\centering
\includegraphics[width=0.45\textwidth]{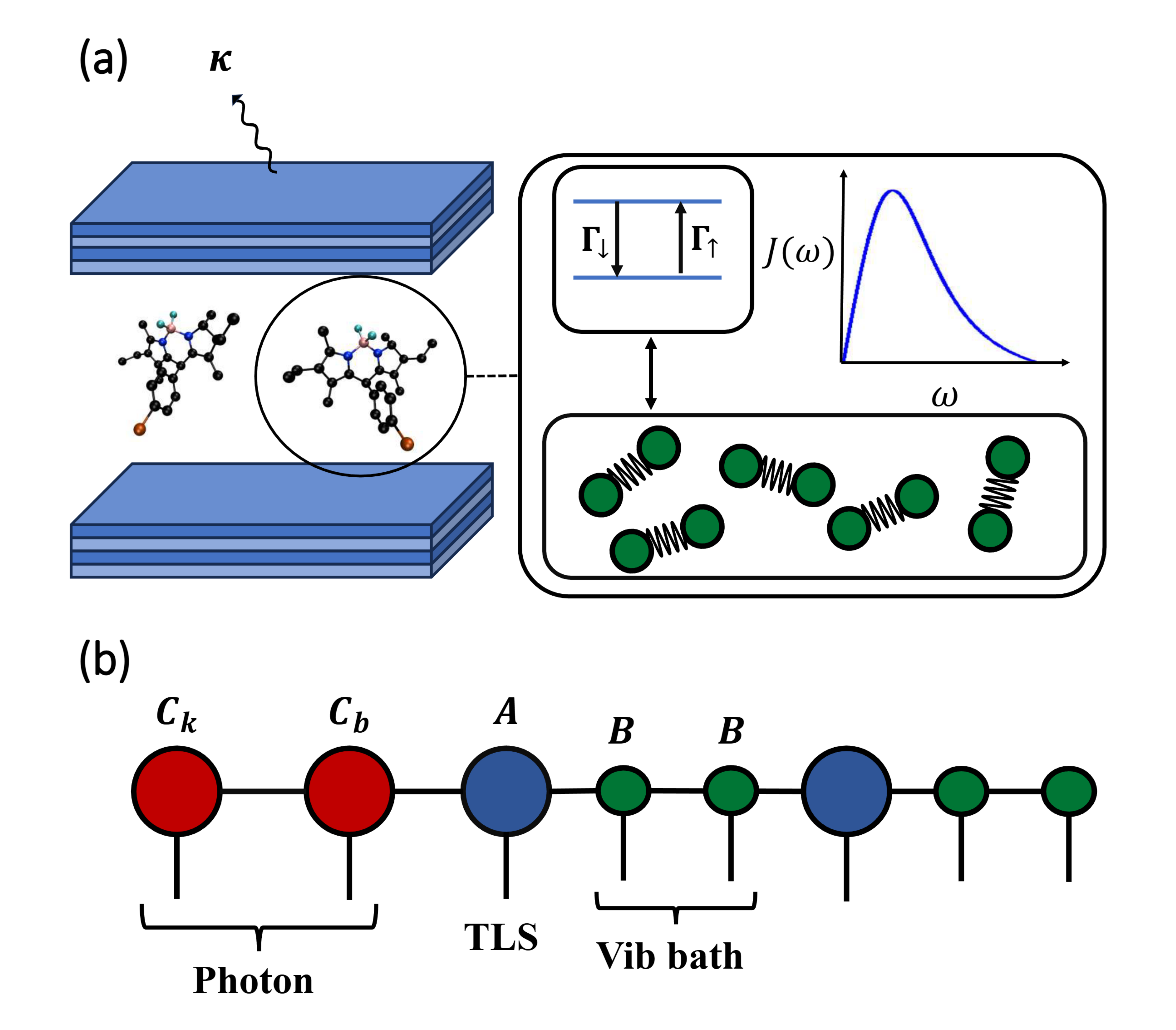}
\vspace{-10pt}
\caption{
(a) Schematic of a molecular ensemble in an optical microcavity. Each molecule is modeled as a TLS coupled to a vibrational environment characterized by spectral density, $J(\omega)$. (b) MPS representation of the ADOs for the HTC model with two TLSs. Red and blue circles represent the cavity and TLS indices, respectively, while green circles track HEOM indices for the two effective vibrational modes.}
\label{fig}
\end{figure}

Matter in polaritons is pumped and lies in a lossy cavity. Hence, one must include incoherent pumping and dissipation. We quantify Markovian cavity loss with rate $\kappa$, and the incoherent pumping, decay, and dephasing of the TLSs with rates $\Gamma_\uparrow$, $\Gamma_\downarrow$, and $\Gamma_\phi$, respectively. Since these processes arise from weak coupling to the external environment compared to the strong light–matter interaction, we capture their contribution with a Lindbladian description~\cite{fowler22}, $\mathcal{R}[\rho] \equiv 2\kappa \mathcal{D}[a] + \sum_{i=1}^N \left( \Gamma_\uparrow \mathcal{D}[\sigma_i^+] + \Gamma_\downarrow \mathcal{D}[\sigma_i^-] +\Gamma_\phi \mathcal{D}[\sigma_i^z]\right)$, where $\mathcal{D}[x] \equiv x\rho x^\dagger - \frac{1}{2} \{ x^\dagger x, \rho \}$. Hence, the equation of motion for the density matrix, $\rho$, in the presence of this pumping and dissipation becomes $\partial_t \rho = -i \mathcal{L} \rho + \mathcal{R}[\rho]$, where, $\mathcal L = [H,...]$ is the Liouville operator for the TC or HTC models.


\textit{Methods} --- We build upon recent algorithmic advances in MPSs \cite{jaschke2018one,finsterholzl2020using,ding2026simulation} to tame the computational complexity of simulating pumped–dissipative HTC models. The central challenge is that each molecule interacts with a vibrational environment whose dynamics lie beyond weak-coupling or Markovian approximations~\cite{thorwart2009enhanced,prior2010efficient}. Although the nonperturbative HEOM formalism can capture these effects, its exponential scaling has so far restricted simulations to the few-emitter regime, far from the thermodynamic limit~\cite{lindoy23}. To overcome this bottleneck, we introduce an MPS representation of the auxiliary density operators (ADOs) appearing in HEOM,
\begin{align}
    \rho_{\mathbf{n}} =& \sum_{\{l\}} 
    C_k^{[0]} C_b^{[1]}A^{[2]}B^{[3]}\cdots B^{[k+2]}
    \cdots \nonumber \\
    &A^{[(N-1)(K+1)+2]}B^{[(N-1)(k+1)+3]}\cdots B^{[N(K+1)+1]}.
\end{align}
where the operators $\rho_{\mathbf{n}}$ denote the hierarchy of auxiliary density operators (ADOs), indexed by the multi-index $\mathbf{n}=\{\{n_{10},n_{11},\ldots\},\ldots\}$.
These ADOs evolve according to
\begin{align}
\label{eq:heom}
    \partial_t \rho_{\bf n}& = -i \mathcal{L} \rho_{\bf n} + \mathcal{R}[\rho_{\bf n}] 
    - \sum_{j,k} n_{jk}\gamma_k\rho_{\bf n} 
    -i\sum_{j}\left[\sigma_j^z,\sum_{k}\rho_{{\bf n}_{jk}^+}\right]  \nonumber \\
    &\quad - i\sum_{j,k} n_{jk}\left(d_k\sigma_j^z\rho_{{\bf n}^-_{jk}}-d_k^*\rho_{{\bf n}^-_{jk}}\sigma_j^z\right),
\end{align}
where the coefficients $d_k$ and $\gamma_k$ arise from the exponential decomposition of the bath correlation function 
$C(t)=\sum_k d_k e^{-\gamma_k t}$~\cite{shi09b,shi09c,liu14}.

The resulting tensor-network structure contains $2+N(K+1)$ sites (see Fig.~\ref{fig}b). Our construction includes ket and bra cavity sites ($C_k$ and $C_b$) and physical TLS sites represented via the Choi transformation ($A$)~\cite{li2023tangent,li2025numerically}, while the additional $NK$ sites encode the HEOM hierarchy within the tensors $B$~\cite{shi18,ke22,guan24}.

Because realistic systems often exhibit chemical heterogeneity, spatial inhomogeneity of the cavity field, and variations in molecular orientations, in our HEOM hybrid, we include disorder in local exciton frequencies, $\omega_0$, and light–matter coupling strengths, $\Omega$. Including frequency disorder is easy within our MPS-HEOM hybrid. For Gaussian-distributed disorder with standard deviation $\sigma$, the bath correlation function acquires an additional constant term, $\tilde{C}(t) = C(t) + \sigma^2$, that introduces an extra component into the multi-index $\mathbf{n}$ of ADOs~\cite{huang2024simulation}. Like frequency disorder, coupling disorder can be mapped onto a simple bath correlation function augmentation, with the bath coupled to the two-body light and matter interaction operator $a^\dagger \sigma_i^- + a \sigma_i^+$(see Supplementary Materials(SM) Sec.~II for details). 

Our MPS–HEOM hybrid reduces the computational cost sufficiently to simulate systems with up to $N\sim100$ TLSs in the intermediate-coupling regime. We employ TDVP time evolution with GPU acceleration~\cite{haegeman16,lubich18}, yielding linear scaling of the computational cost with system size (see SM~Sec.~I). This efficiency enables simulations far beyond the reach of previous non-Markovian approaches. Importantly, our MPS formulation is not restricted to the single-excitation manifold and can treat multiphoton states of the cavity.


\textit{Results and discussion} --- Given the discrepancy between system sizes in experiment and theory, we investigate polariton dynamics using our hybrid MPS--HEOM framework to determine the thermodynamic threshold system size, $N_T$, relevant for polaritonic systems. We determine $N_T$ by evaluating the time-normalized root mean square error (RMSE) of the dynamics between consecutive TLS numbers (SM~Eq.~S6), with the convergence threshold set to $10^{-4}$. Identifying $N_T$ is also necessary for further theoretical developments. For instance, in the thermodynamic limit, a TLS bath can be mapped onto a harmonic oscillator analogue~\cite{makri95b,suarez91a,wang2012dynamics,ying2024spin}, offering a path for simplifying its theoretical treatment. Further, knowing $N_T$ informs \textit{ab initio} simulations of molecular polaritons as to how many molecules are required to simulate collective polaritonic behavior \cite{li2022qm, welman2025light}.  However, how large a polaritonic system must be to effectively reach this limit in the presence of static disorder (SD) and dynamic disorder (DD) remains unexplored. Here, we address this question by explicitly determining $N_T$ as a function of disorder strength.

We begin by analyzing the experimentally measurable average photon number $\langle a^\dagger a \rangle$, which characterizes photon dynamics and is widely used to identify lasing and superradiant phase transitions~\cite{kirton17, kirton18, li25}. Throughout this work, we initialize the system in the state 
$|1\rangle\langle1|\otimes \rho_g^{\rm TLS}\otimes e^{-\beta\sum_i H_B^i}$, corresponding to a single photon excitation in the cavity, all TLSs prepared in their ground state, and thermal equilibrium for the vibrational baths. We set the system parameters to align with experimental studies of organic polaritons like BODIPY-Br~\cite{grant2016efficient,cookson2017yellow,dietrich2016exciton}, with $\omega_c = \omega_0 = 2.0$ and $\kappa = \Gamma_\downarrow = 0.02$, corresponding to a cavity quality factor of $Q = 100$. In systems with dynamic disorder, TLSs couple with local phonon baths that modulate their energies. In the limit of a slow vibrational bath, the variance of the energy fluctuations is related to the reorganization energy $\lambda$ and the temperature $T$ through the fluctuation--dissipation relation~\cite{nitzan01}. Specifically, we use the root-mean-square amplitude of the energy fluctuations, $\sigma_{\mathrm{dyn}} = \sqrt{\langle \delta\omega^2 \rangle}= \sqrt{2\lambda k_B T}$, to quantify the strength of dynamic disorder, and compare to cases with SD. 

\begin{figure}[t]
\centering
\includegraphics[width=8.5cm]{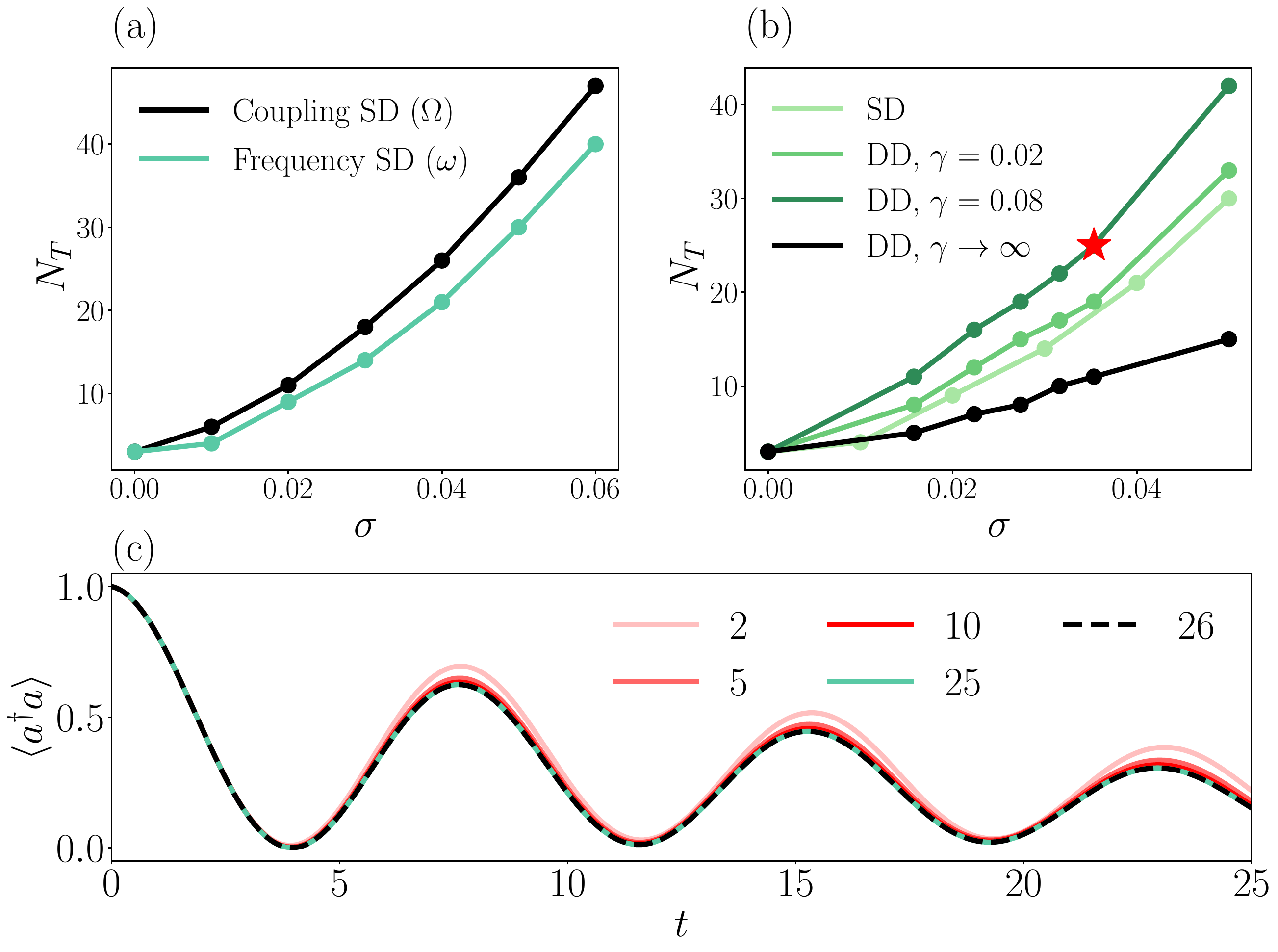}
\vspace{-8pt}
\caption{
(a) Threshold system size, $N_T$, as a function of disorder strength $\sigma$ for frequency versus light–matter coupling disorder.
(b) Comparison of $N_T$ obtained for static (SD) and dynamic (DD) frequency disorder for three bath characteristic frequency, $\gamma=0.02$, $\gamma=0.04$ and $\gamma \to \infty$.
(c) Time evolution of the average photon number $\langle a^\dagger a\rangle$ for different TLS numbers $N$ at fixed $\eta = 0.05$($\sigma = 0.0354$), corresponding to the red star in panel (b).
}
\label{fig2}
\end{figure}

Figure~\ref{fig2} analyzes how frequency versus coupling disorder affect convergence to the thermodynamic limit and how the latter depends on the timescale of frequency disorder. Interestingly, Fig.~\ref{fig2}(a) shows that $N_T = 3$ in the absence of any disorder ($\sigma=0$), reflecting the highly collective nature of the idealized system---an insight exploited by the CUT-E method ~\cite{perez23, perez25}. Figure~\ref{fig2}(a) further demonstrates that SD in the light–matter coupling, $\Omega + \delta \Omega_i$, requires larger values of $N_T$ than frequency disorder, $\omega + \delta \omega_i$. Figure~\ref{fig2}(b) examines the convergence to the thermodynamic limit as a function of the speed of vibrational equilibration, $\gamma$, ranging from the inhomogeneous (static, $\gamma = 0$) to the homogeneous (Markovian, $\gamma \rightarrow \infty$) limits. This rate of convergence becomes progressively slower (i.e., requires increasingly larger $N_T$) as the bath becomes faster from the static-disorder limit into the weakly Markovian regime, with the inhomogeneous limit providing a lower bound for the required system size, $N_T$, in this parameter range. However, this trend does not persist indefinitely. As $\gamma \rightarrow \infty$, $N_T$ remains at a value even lower than that required by the inhomogeneous limit. Consequently, the dependence of the convergence rate on $\gamma$ is non-monotonic and exhibits turnover behavior, reminiscent of a Kramers turnover~\cite{kramers40,hanggi90}. 

To uncover the microscopic origin of this turnover behavior, we analyze how disorder modifies the participation of non-collective degrees of freedom in the dynamics. We track the total exciton population $P_{\rm tot}=\sum_i\langle\sigma_i^+\sigma_i^-\rangle$, which reflects the efficiency of collective light--matter exchange, together with the bright state and the dark state populations: $P_{\rm bright}=1/N\sum_{i,j}\langle\sigma_i^+\sigma_j^-\rangle$ and $P_{\rm dark}=P_{\rm tot}-P_{\rm bright}$. While $P_{\rm tot}$ determines how efficiently a cavity excitation is transferred into the matter subsystem, $P_{\rm dark}$ reveals whether population accumulates in non-collective manifolds. Together, these observables allow us to assess how collective exchange is suppressed and how non-collective degrees of freedom become dynamically involved.

The roles of frequency and coupling disorder can be understood from their qualitatively different impacts on collective light–matter behavior. At the microscopic level, these differences originate from how each type of disorder restructures and activates the dark state manifold. In the absence of coupling disorder, the light–matter Hamiltonian can be block-diagonalized into a bright subspace that couples collectively to the cavity mode and a dark subspace that is completely decoupled from the photonic degree of freedom (SM~Fig.~S3). While frequency disorder couples the bright and dark manifolds through the matter Hamiltonian, facilitating population transfer from the bright state into dark states, the dark states remain uncoupled from the cavity mode. By contrast, coupling disorder explicitly breaks the collective symmetry of the light–matter interaction, allowing the cavity mode to couple directly to both bright and dark states. The corresponding derivations are provided in the SM. In this case, the notion of a dark state manifold breaks down, and dark states become optically active gray states~\cite{sauerwein2023engineering}. 

\begin{figure}[t]
\centering
\includegraphics[width=8.70cm]{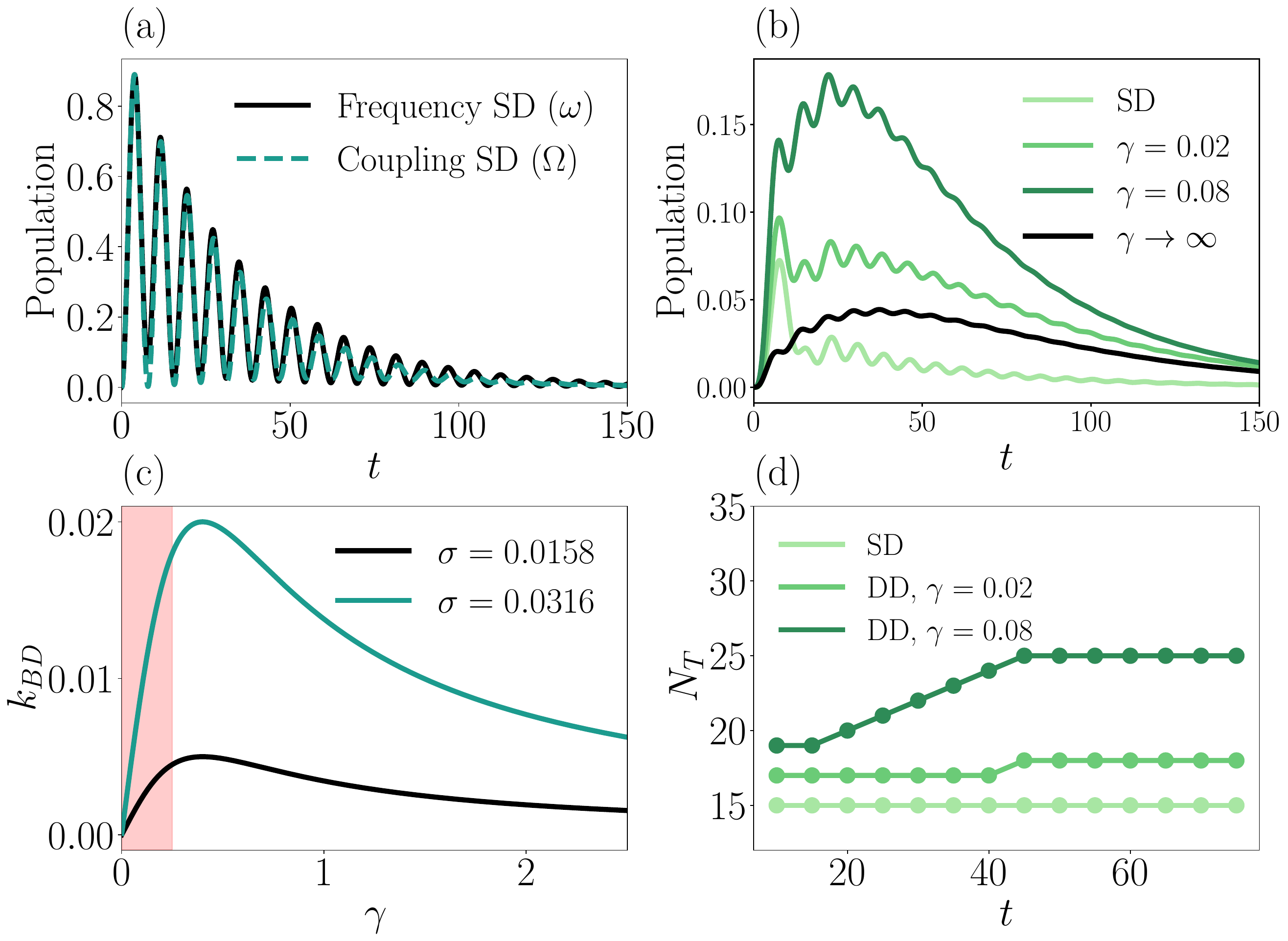}
\vspace{-20pt}
\caption{
(a) Evolution of the total exciton population for SD in the coupling versus frequency.
(b) Evolution of the dark state population at $\sigma = 0.0316$ for static disorder and dynamic disorder with different bath characteristic frequencies ($\gamma$).
(c)The bright--to--dark state transfer rate versus $\gamma$ for different disorder strengths calculated within perturbative rate theory. The red-shaded region highlights the breakdown of the perturbative theory.
(d) Time dependence of the threshold system size $N_T$ for SD and DD.
}
\vspace{-10pt}
\label{fig3}
\end{figure}

These structural differences manifest in the dynamics. Figure~\ref{fig3}(a) compares the evolution of the total exciton population $P_{\rm tot}(t)$ under frequency versus coupling disorder. While both cases exhibit oscillatory light–matter exchange, coupling disorder leads to faster damping of the oscillations. This indicates a reduction of coherent collective exchange and faster redirection of excitation into dark and gray states. In fact, recent studies have shown that disorder in the TC model can activate non-collective states, causing anomalous transport and the absence of thermalization~\cite{mattiotti2024multifractality}. This observation has an important consequence: as more excitation flows into non-collective (dark and gray) states, larger system sizes are necessary to reach the thermodynamic limit.

Building on this insight, the observed trends in $N_T$ can be attributed to how disorder suppresses collective behavior through the dynamic involvement of non-collective states. In the presence of frequency disorder, dark state population grows with decreasing collective behavior. As Fig.~\ref{fig3}(b) shows, static disorder leads to a small occupation of dark states, indicating that only a small amount of excitation leaks out of the collective bright sector. As a result, collective light–matter behavior is only weakly suppressed, and convergence to the thermodynamic limit requires a small $N_T$. By contrast, in the presence of dynamic disorder, the dark state population exhibits a turnover behavior, increasing from the small-$\gamma$ regime, and decreasing again in the large-$\gamma$ limit. A simple perturbative treatment explains this turnover behavior. In SM~Sec.~V, we show that dark state population governed by the bright--to--dark transfer rate is primarily determined by the spectral weight of the noise spectrum at the Rabi splitting. Figure~\ref{fig3}(c) shows that the perturbative theory reproduces the same turnover behavior as observed in the numerically exact results and predicts disorder-dependent $\gamma$ that optimizes flow into the dark manifold. As expected based on previous perturbative studies~\cite{ishizaki09,montoya2015extending}, such approximations underestimates the transfer rate in the small-$\gamma$ regime relative to our exact dynamics. Nevertheless, this explanation provides insight on the importance of polariton--vibrational timescale matching.

This interpretation is further supported by the dependence of $N_T$ on the total simulation time shown in Fig.~\ref{fig3}(c). In the less non-Markovian regime ($\gamma = 0.08$), $N_T$ initially increases with the observation time, reaches a maximum around $t \approx 40$, and subsequently decreases before saturating at a finite plateau. This behavior closely mirrors the evolution of the dark state population, indicating that the transient growth of $N_T$ is associated with enhanced population transfer into the dark manifold, while its later saturation reflects the stabilization of the dark state occupation. By contrast, in the more non-Markovian regime (static disorder and dynamic disorder with $\gamma = 0.02$), $N_T$ exhibits little time dependence and remains essentially constant, consistent with the much smaller and weakly time-dependent dark state population in these cases. These results demonstrate that the behavior of $N_T$ is governed by the dynamics of dark state population and the suppression of collective behavior.

\textit{Conclusions} --- We have developed a hybrid MPS--HEOM framework that enables numerically exact simulations of polariton dynamics from a few emitters to the thermodynamic limit under both static and dynamic disorder. For the first time, this method allows us to quantitatively answer the long-standing question of what is the minimum system size required to reach the thermodynamic limit in collective polaritonic systems. By systematically analyzing the convergence scale $N_T$ with static and dynamic disorder of the chromophore energies and light-matter couplings, we discover that the approach to the thermodynamic limit becomes more demanding in the presence of dynamic rather than static disorder, and exhibits a non-monotonic dependence on bath Markovianity, first increasing and then decreasing as the bath becomes more Markovian. Our method can also be used to guide minimal system sizes for \textit{ab initio} simulations to study collective polariton behavior for systems with many molecules~\cite{li2022qm,  welman2025light, wickramasinghe2025fly}.

We have identified the microscopic origin of the observation that disorder, and especially dynamic disorder, suppresses collective light–matter dynamics by activating non-collective degrees of freedom. This provides the key link between disorder-induced breakdown of collective behavior and convergence to the thermodynamic limit. Our results further demonstrate that phonon timescales and the extent of non-Markovianity play a decisive role in regulating bright–dark transfer, resulting in a strong and non-monotonic dependence of the threshold system size $N_T$ on the bath correlation time that correlates directly to a Kramers turnover in rate of bright-to-dark energy flow. Beyond open quantum many-body dynamics, our results provide new physical insight into polariton physics by identifying the conditions under which dark states remain dynamically decoupled and those under which they become actively involved in polariton-mediated processes. More broadly, this work establishes a general framework for understanding how colored environments mold collective light–matter dynamics and control the emergence of thermodynamic behavior in strongly coupled systems.

\section*{acknowledgement}

A.~M.~C., T.~L., and P.~V.~were supported by an Early Career Award in the CPIMS program in the Chemical Sciences, Geosciences, and Biosciences Division of the Office of Basic Energy Sciences of the U.S.~Department of Energy under Award DE-SC0024154. A.M.C.~also acknowledges the support from a David and Lucile Packard Fellowship for Science and Engineering. QS thanks the support from the National Natural Science
Foundation of China (NSFC) (Grant No. 22433006). We thank Juan B.~Perez-Sanchez, Joel Yuen-Zhou, and Hsing-Ta Chen for comments on the manuscript. This work utilized the Alpine high-performance computing resource at the University of Colorado Boulder. Alpine is jointly funded by the University of Colorado Boulder, the University of Colorado Anschutz, Colorado State University, and the National Science Foundation (Award No. 2201538).


\bibliography{quantum}

\pagebreak

\end{document}


\title{Supplemental Material for\\For molecular polaritons, disorder and phonon timescales control the activation of dark states in the thermodynamic limit}

\author{Tianchu Li}
\affiliation{Department of Chemistry, University of Colorado Boulder, Boulder, Colorado 80309, USA\looseness=-1}
\author{Pranay Venkatesh}
\affiliation{Department of Chemistry, University of Colorado Boulder, Boulder, Colorado 80309, USA\looseness=-1}
\author{Qiang Shi}\email{qshi@iccas.ac.cn}
\affiliation{Beijing National Laboratory for
Molecular Sciences, State Key Laboratory for Structural Chemistry of Unstable and Stable Species, Institute of Chemistry, Chinese
Academy of Sciences, Zhongguancun, Beijing 100190, China}
\affiliation{University of Chinese Academy of Sciences,
Beijing 100049, China}

\author{Andr\'es Montoya-Castillo}
\email{Andres.MontoyaCastillo@colorado.edu}
\affiliation{Department of Chemistry, University of Colorado Boulder, Boulder, Colorado 80309, USA\looseness=-1}
\maketitle

\section{Computational details for hybrid HEOM MPS method}

\begin{figure}[htbp]
\renewcommand{\figurename}{FIG.}              
\renewcommand{\thefigure}{S\arabic{figure}}
\centering
\includegraphics[width=15cm]{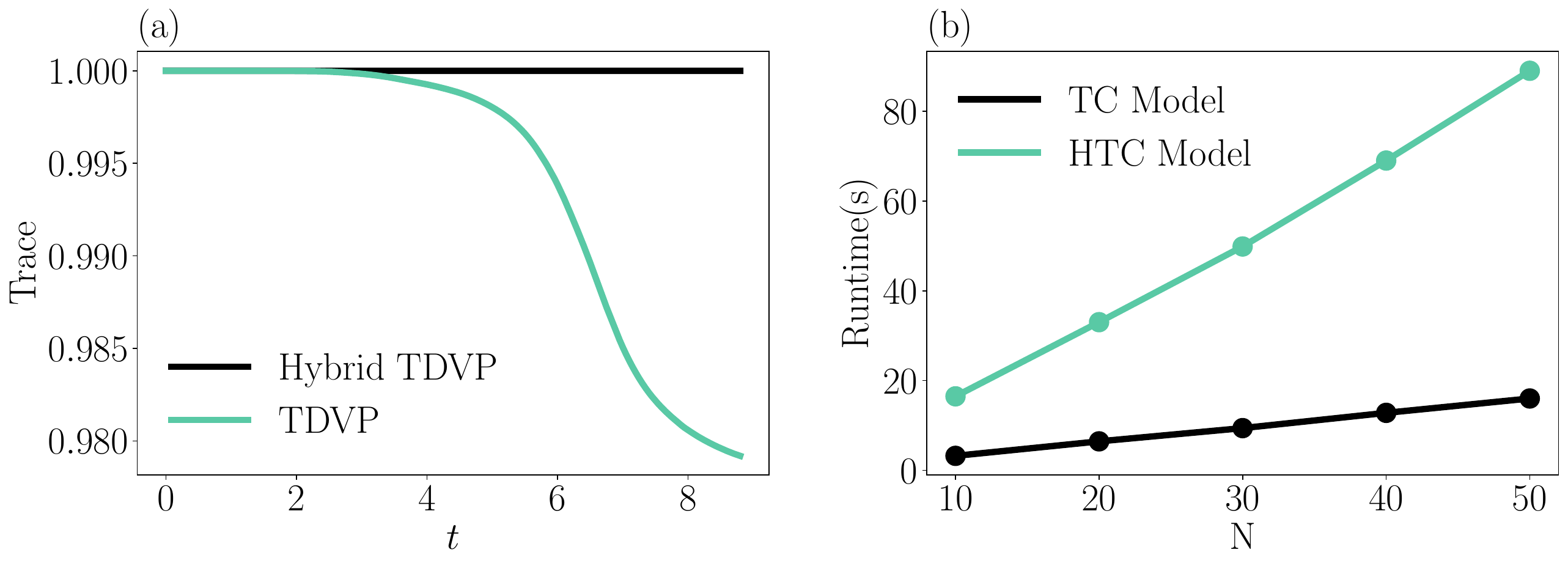}
\vspace{1em}
\caption{(a) Evolution of the trace of the photon density operator for the TC model with $N = 80$. The bond dimension for TDVP is set as 200. (b) Computational time per 100 time steps as a function of the number of TLSs $N$ for both the TC and HTC models.
}
\label{figs1}
\end{figure}

One of the main challenges of combining matrix product states (MPS) with the time-dependent variational principle (TDVP) is maintaining trace conservation, since the MPS is usually propagated within a fixed-rank manifold~\cite{lubich15}.
To tackle this problem, we adopt a hybrid propagation strategy in which the fourth-order Runge–Kutta (RK4) method is used for the first 10 time steps, followed by TDVP evolution for the remaining dynamics~\cite{haegeman16,ceruti2022unconventional}. As shown in Fig.~\ref{figs1}(a), this hybrid scheme significantly improves trace conservation compared with standard TDVP propagation.

We treat the cavity mode using the potential-optimized discrete variable representation (PO-DVR)~\cite{colbert92}, with the number of basis functions denoted by $N_{\mathrm{DVR}}$, and implement the propagation with GPU acceleration~\cite{li2025numerically}. In all simulations, we set $N_{\mathrm{DVR}}=2$, which is sufficient within the single-excitation manifold considered here.
The GPU implementation provides an approximately $60\times$ speedup over the CPU version. For $N=10$ TLSs, the TC model requires $\sim0.3$ seconds per timestep, while the HTC model requires $\sim2$ seconds per timestep on an NVIDIA A800 GPU.  The computational cost per timestep scales linearly with the number of TLSs, as shown in Fig.~\ref{figs1}(b). In all TDVP simulations, the bond dimension is fixed to 200 for the Tavis–Cummings (TC) model and 250 for the Holstein–Tavis–Cummings (HTC) model.

\begin{figure}[htbp]
\renewcommand{\figurename}{FIG.}               
\renewcommand{\thefigure}{S\arabic{figure}}
\centering
\includegraphics[width=15cm]{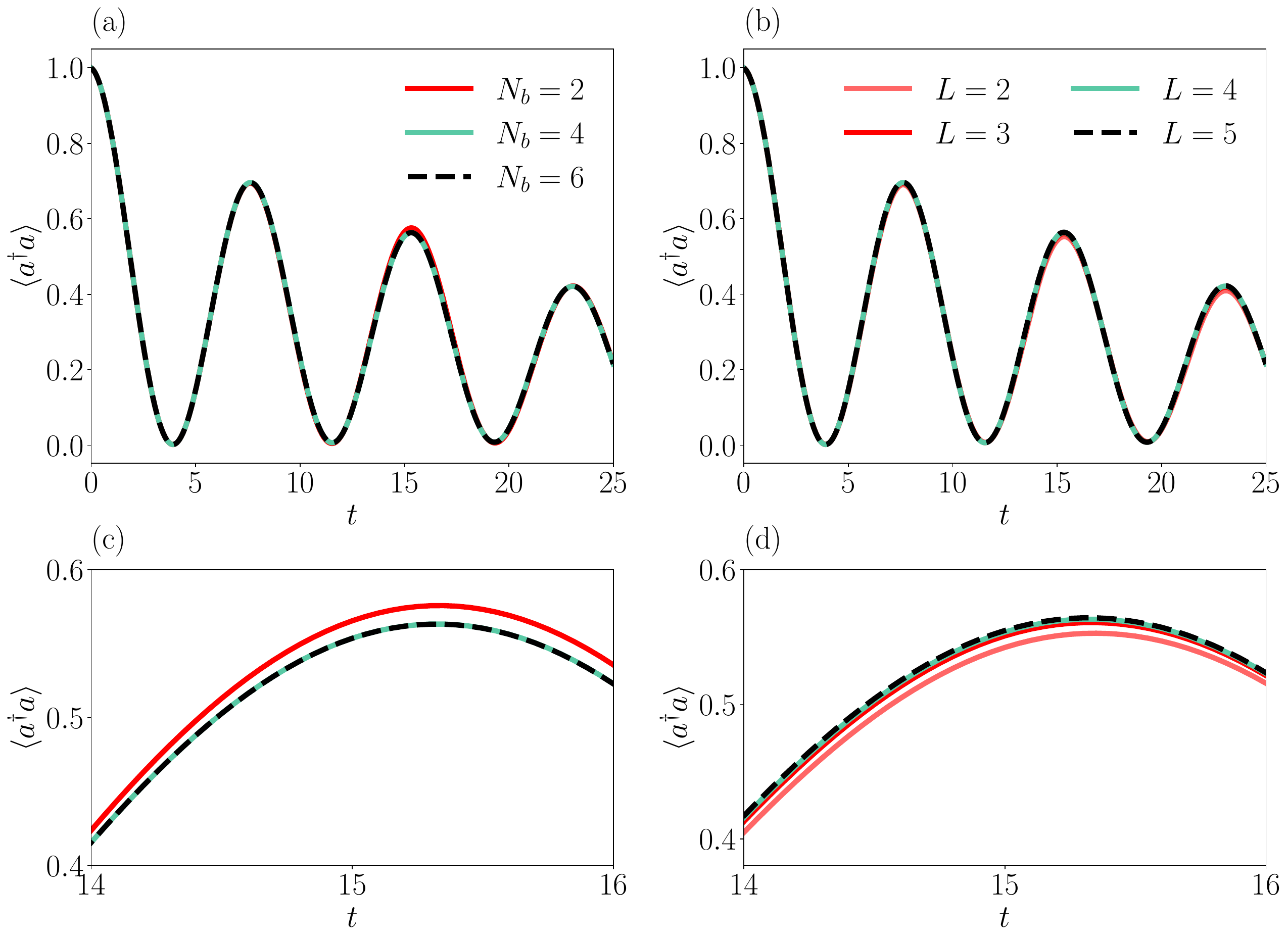}
\vspace{1em}
\caption{Evolution of the average photon number, $\langle a^\dagger a\rangle$, showing convergence with respect to (a) the hierarchical depth $N_b$ and (b) the number of exponential terms $L$  in dynamic disorder case. Panel (c) is an inset of panel (a) and  Panel (d) is an inset of panel (b). The parameters are $\eta = 0.05$, $\gamma = 0.08$ and $N = 25$.}
\label{figs2}
\end{figure}

For the hybrid HEOM–MPS approach, the HEOM truncation parameters must be converged. We employ the $N$-truncation at hierarchical depth $N_b$~\cite{li22}, setting $\rho_{\mathbf n}=0$ whenever $n_i>N_b$. The bath correlation function is represented via an exponential decomposition truncated to $L$ terms~\cite{liu14}. In the dynamic disorder case, Figs.~\ref{figs2}(a) and \ref{figs2}(b) show the convergence of the average photon dynamics with respect to $N_b$ and $L$ for $\eta=0.05$, $\gamma=0.08$, and $N=25$, yielding converged results at $N_b=4$ and $L=5$. This parameter set corresponds to the most numerically demanding regime considered in this work; all simulations in the main text therefore employ these truncation parameters.

\begin{figure}[htbp]
\renewcommand{\figurename}{FIG.}               
\renewcommand{\thefigure}{S\arabic{figure}}
\centering
\includegraphics[width=15cm]{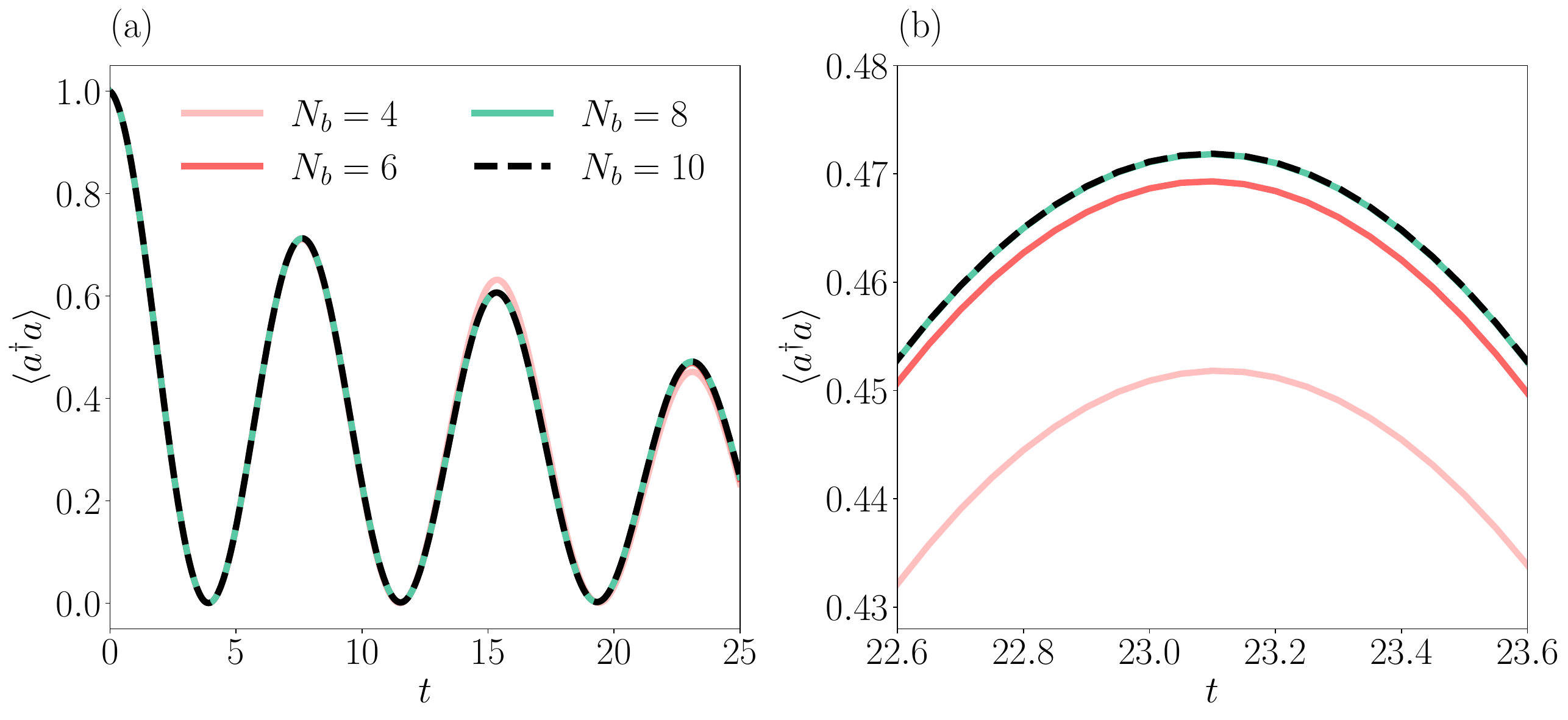}
\vspace{1em}
\caption{(a) Evolution of the average photon number, $\langle a^\dagger a\rangle$, showing convergence with respect to  the hierarchical depth $N_b$ in static disorder case. Panel (b) is an inset of panel (a). The parameters are $\sigma = 0.035$ ($\eta=0.05$) and $N = 25$.}
\label{figs3}
\end{figure}

In the static disorder case, to gain a better numerical stability of HEOM, one need to scale the auxiliary density operators(ADOs) to~\cite{shi09b}
\begin{align}
\tilde\rho_{\mathbf n} = \left(\prod_k \sigma^{2n_k}n_k!\right)^{-\frac{1}{2}}\rho_{\mathbf n},
\end{align}
and the correpsponding HEOM can be written as:
\begin{align}
    \partial_t\tilde\rho_{\mathbf n} =
    -i\mathcal L_{\rm TC}\tilde\rho_{\mathbf n}
    -i\sum_j \sigma\sqrt{n_j+1}\left[\sigma^z_j,\tilde\rho_{\mathbf n_j^+}\right]
    -i\sum_j \sigma \sqrt{n_j} \left[\sigma^z_j,\tilde\rho_{\mathbf n_j^-}\right].
\end{align}
Figure~\ref{figs3} shows the convergence of the average photon dynamics with respect to $N_b$ in the static-disorder case at the same disorder strength as in Fig.~\ref{figs2}. A larger hierarchical depth ($N_b = 8$) is required to reach convergence than in the dynamic-disorder case. This result is consistent with statement in Ref.~\onlinecite{perez2024collective}, where static disorder requires a larger computational resources than dynamic disorder in the thermodynamic limit.

\section{Hierarchical equations of motion for TC model with coupling static disorder}

For the TC model with static disorder in the light--matter coupling, we assume the disorder follows a Gaussian distribution with standard deviation $\sigma$. 
A Gaussian static disorder can be mapped onto an effective harmonic oscillator bath because a Gaussian-distributed random variable is fully characterized by its second cumulant and produces the same influence functional as a bosonic bath with a delta-correlated correlation function. 

The resulting Hamiltonian can therefore be written as
\begin{align}
\tilde H = H_{\rm TC} + \sum_\alpha \left[
c_\alpha(b^+_\alpha+b_\alpha) \sum_i \left(a^\dagger\sigma_i^- + a\sigma_i^+\right)
+ \omega_\alpha b_\alpha^\dagger b_\alpha
\right],
\end{align}
where $b_\alpha^\dagger$ and $b_\alpha$ denote the creation and annihilation operators of the effective bath modes. The effective harmonic oscillator bath follows Debye-Drude spectral density with $\gamma=0$. And the corresponding bath correlation function reduces to a time-independent form $C(t)=\sigma^2$.

The HEOM can therefore be expressed as
\begin{align}
\partial_t\rho_{\mathbf n} =
-i\mathcal L_{\rm TC}\rho_{\mathbf n}
-i\sum_j \left[g_j,\rho_{\mathbf n_j^+}\right]
-i\sum_j \sigma^2 n_j \left[g_j,\rho_{\mathbf n_j^-}\right],
\end{align}
where
\begin{align}
g_j = a^\dagger\sigma_j^- + a\sigma_j^+
\end{align}
is the two-body light and matter interaction operator.

\section{Convergence Analysis of dynamics Based on the Time-Normalized root mean square error}
In this work, we determined the convergence by calculating the time-normalized root mean square error between two simulations $P_1$ and $P_2$, where the error is defined as:
\begin{align}
\mathrm{RMSE} = \sqrt{\frac{1}{t_{\rm max}}\int_0^{t_{\rm max}}dt\left[P_1(t)-P_2(t)\right]^2}.
\end{align}
Here, to avoid artificial dilution of the error by the long-time tail where the dynamics has largely decayed to $0$, we evaluate the time-normalized RMSE over a finite observation window $[0,t_{\max}]$. In this work, we set $t_{\max}$ as the time at which the population dynamics decays to $10^{-3}$.

\section{Dark state population under different dissipative channel}
\begin{figure}[htbp]
\renewcommand{\figurename}{FIG. }                 
\renewcommand{\thefigure}{S\arabic{figure}}
\centering
\includegraphics[width=16cm]{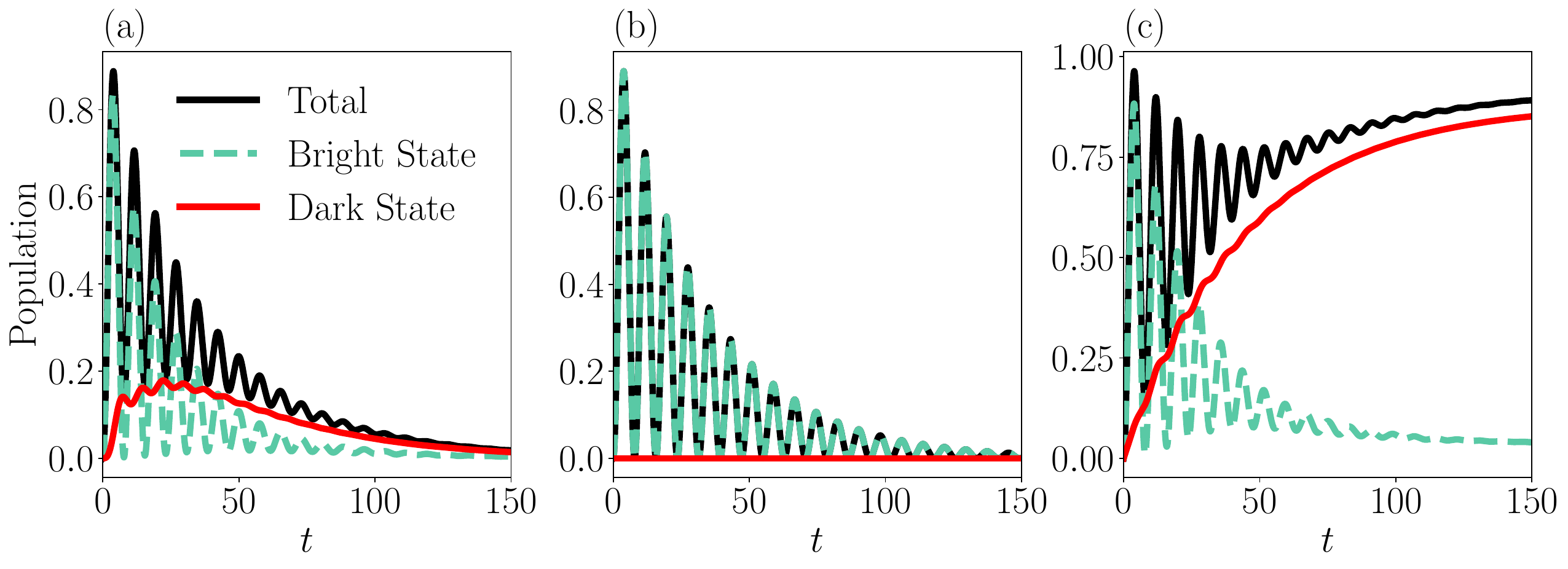}
\vspace{1em}
\caption{Evolution of the total, bright, and dark state populations for TC model with (a) dynamic disorder ($\gamma=0.08$), (b) spontaneous emission Lindbladian dissipation ($\Gamma_\downarrow$), and (c) an incoherent pumping Lindbladian contribution ($\Gamma_\uparrow$).}
\label{figs4}
\end{figure}

In the presence of frequency disorder, a finite occupation of the dark states emerges, as shown in Fig.~\ref{figs4}. This occurs because the dephasing-type disorder breaks the permutation symmetry of the system and gradually redistributes population from the collective bright state into the dark manifold.

In the absence of both disorder and incoherent pumping, no population of dark states is observed, as shown in Fig.~\ref{figs4}(b). This can be understood from symmetry considerations: the decoherence Lindbladian preserves the collective symmetry of the Tavis--Cummings Hamiltonian and therefore only mediates population transfer between the bright state and the ground state, without coupling to the dark manifold.

When incoherent pumping is included, however, a finite dark state population emerges, as shown in Fig.~\ref{figs4}(c). This is because the incoherent pumping Lindbladian injects excitations locally and randomly into individual emitters, thereby breaking the collective selection rule and populating both bright and dark states.

\section{Dark state population under disorder}

Frequency and coupling disorder populate dark states through qualitatively different mechanisms. Before considering the effect of disorder, we employ the disorder-free case to introduce the concepts of bright and dark states. In this limit, the collective bright state in the single-excitation manifold is defined as
\begin{align}
    |B\rangle = \frac{1}{\sqrt{N}} \sum_{i=1}^{N} |i\rangle,
\end{align}
where $|i\rangle \equiv |0,\ldots,1_i,\ldots,0\rangle$ denotes a single excitation localized on the $i$th TLS. The corresponding single-excitation basis including the cavity mode is defined as:
\begin{align}
|1_c,G\rangle \equiv |1\rangle_{\rm cav} \otimes |g,\cdots,g\rangle, \qquad 
|0_c,i\rangle \equiv |0\rangle_{\rm cav} \otimes |i\rangle.
\end{align}
Dark states are defined as matter states orthogonal to $|B\rangle$
\begin{align}
    |D_\mu\rangle = \sum_{i=1}^{N} \tilde d_i^{(\mu)} |i\rangle,
\end{align}
with coefficients satisfying the constraint
\begin{align}
    \sum_{i=1}^{N}\tilde d_i^{(\mu)} = 0.
\end{align}

We can now consider the case of frequency disorder in the matter Hamiltonian,
\begin{align}
    H_{\rm TLS} = \sum_{i=1}^{N} (\omega_0 + \delta\omega_i)\sigma_i^z,
\end{align}
where $\delta\omega_i$ are random frequency shifts. In the $\{|B\rangle, |D_\mu\rangle\}$ basis, this term generates off-diagonal matrix elements,
\begin{align}
\label{eq:bdcoupling}
\langle D_\mu | H_{\rm TLS} | B \rangle = \frac{1}{\sqrt{N}} \sum_{i=1}^{N} \delta\omega_i \tilde d_i^{(\mu)},
\end{align}
which are generically nonzero. As a result, frequency disorder induces coupling between the bright and dark manifolds through the matter Hamiltonian, enabling population transfer from the bright state into dark states. By contrast, the light--matter interaction retains its collective form,
\begin{align}
H_{\text{int}} = \frac{\Omega}{\sqrt{N}} \sum_{i=1}^{N} \left(a^\dagger \sigma_i^- + a \sigma_i^+\right),
\end{align}
and therefore the cavity dark state matrix element is zero,
\begin{align}
\langle 1_c,G| H_{\rm int} |0_c, D_\mu\rangle = 0,
\end{align}
indicating that dark states remain strictly uncoupled from the cavity mode in the presence of frequency disorder alone.

We next consider coupling disorder, where the light--matter interaction takes the form,
\begin{align}
H_{\text{int}} = \sum_{i=1}^{N} \frac{\Omega_i}{\sqrt{N}} \left(a^\dagger \sigma_i^- + a \sigma_i^+\right),
\end{align}
with $\Omega_i = \Omega + \delta \Omega_i$. The cavity dark-state matrix element in the single-excitation manifold is then given by
\begin{align}
\langle 1_c,G| H_{\rm int} |0_c, D_\mu\rangle
= \frac{1}{\sqrt{N}}\sum_{i=1}^N \delta\Omega_i\, \tilde d_i^{(\mu)},
\end{align}
which is generically nonzero. Thus, coupling disorder directly activates dark states optically by inducing a finite coupling between the cavity mode and the dark manifold. By contrast, the matter Hamiltonian remains diagonal in the bright--dark basis, so that
\begin{align}
\langle B|H_{\rm TLS}|D_\mu\rangle = 0,
\end{align}
indicating that coupling disorder alone does not induce bright--dark mixing within the matter sector.

\section{Fermi's golden rule for bright--dark energy transfer}

In the presence of frequency disorder, the bright and dark states are coupled through the disorder term in Eq.~\eqref{eq:bdcoupling}. 
Since the disorder is weak in the parameter regimes considered here, we can describe the bright--dark population transfer within a second-order perturbative Fermi's golden rule (FGR) framework.

In the interaction picture with respect to the disorder-free TC Hamiltonian and free phonon environment, phonon-induced disorder couples bright and dark states. This bright--dark coupling reads
\begin{align}
    H_{\rm BD}(t) =\sum_{\mu} \Big(\xi_{\mu}(t)|B\rangle\langle D_{\mu}|+\xi_{\mu}^{\dagger}(t)|D_{\mu}\rangle\langle B|\Big),
\end{align}
where the disorder-induced coupling matrix element is
\begin{align}
    \xi_{\mu}(t)&\equiv \langle B|H_{\rm dis}(t)|D_{\mu}\rangle \nonumber\\
    &= e^{i\Delta_{{\rm BD}_{\mu}}t}\left(\sum_{i=1}^{N} \tilde d_{B,i}^{*}\,\tilde d_{i}
    ^{(\mu)}\,\delta\omega_i(t)\right)\equiv e^{i\Delta_{{\rm BD}_{\mu}}t}\tilde\xi_\mu(t). 
\end{align}
Here, $\Delta_{{\rm BD}_\mu}\equiv E_{D_\mu}-E_B$ is the bright--dark energy splitting, and $H_{\rm dis} = \sum_i^N\delta\omega_i|i\rangle\langle i|$, where $\delta \omega_i = \sum_{\alpha} c_{i,\alpha}(\hat{b}_{i,\alpha}^{\dagger} + \hat{b}_{i,\alpha})$. $\tilde d_{B,i}\equiv \langle i|B\rangle$ and $\tilde d_i^{(\mu)}\equiv \langle i|D_\mu\rangle$ denote the excitonic amplitudes of the bright and dark states on site $i$, respectively. Within second-order perturbation theory, the bright-to-dark transition rate is given by
\begin{align}
    k_{B\to D_{\mu}} 
    =
    2\,\mathrm{Re}\int_{0}^{\infty} dt\;
    \left\langle \tilde\xi_{\mu}(t)\,\tilde\xi^\dagger_{\mu}(0)\right\rangle
    e^{i\Delta_{{\rm BD}_\mu}t}.
\label{eq:kernel_fgr_clean}
\end{align}

Defining the phonon bath correlation function,
\begin{align}
C_{\mu}(t)\equiv \left\langle\tilde\xi_{\mu}(t)\tilde\xi^\dagger_{\mu}(0)\right\rangle,
\end{align}
and the corresponding noise spectrum
\begin{align}
\label{eq:ns}
S_{\mu}(\omega)\equiv 
2\,\mathrm{Re}\int_{0}^{\infty} dt\, 
C_{\mu}(t)\,e^{i\omega t},
\end{align}
the rate assumes the compact FGR form
\begin{align}
k_{B\to D_{\mu}} = S_{\mu}(\Delta_{{\rm BD}_\mu}).
\label{eq:fgr_rate_clean}
\end{align}
In the polaritonic regime considered here, the splitting $\Delta_{{\rm BD}_\mu}$ is set by the collective Rabi splitting, 
$\Delta_{{\rm BD}_\mu}\approx\Omega_R \equiv\sqrt{\Omega^2+(\omega_c-\omega_0)^2}$. For a bath with spectral density $J(\omega)$, the noise spectrum entering Eq.~\eqref{eq:fgr_rate_clean} can be written as
\begin{align}
    S_{\mu}(\omega)
    =
    \left(\sum_{i=1}^{N} |\tilde d_{B,i}|^2 |\tilde d_{i}^{(\mu)}|^2\right)
    J(\omega)\coth\!\left(\frac{\beta\omega}{2}\right),
\end{align}
where the state-dependent prefactor arises from the overlap structure contained in $\xi_\mu(t)$. Summing over the dark manifold yields the total transfer rate from the bright state to dark states,
\begin{align}
k_{B\to D}=\sum_{\mu} k_{B\to D_{\mu}}=\sum_{\mu}
\left(\sum_{i=1}^{N} |\tilde d_{B,i}|^2 |\tilde d_{i}^{(\mu)}|^2\right)
J(\Omega_R)\coth\!\left(\frac{\beta\Omega_R}{2}\right).
\label{eq:fgr_total_clean}
\end{align}

For the symmetric bright state, $\tilde d_{B,i}=1/\sqrt{N}$. Together with the normalization condition of the dark states, $\sum_i|\tilde d_i^{(\mu)}|^2 = 1$, the noise spectrum in Eq.~\eqref{eq:ns} reduces to
\begin{align}
S_\mu(\omega)=\frac{1}{N}J(\omega)\coth\!\left(\frac{\beta\omega}{2}\right),
\end{align}
implying identical bright-to-dark transfer rates for all dark states. 
Summing over the $N-1$ dark states therefore yields the total transfer rate
\begin{align}
k_{B\to D} = \frac{N-1}{N}\,J(\Omega_R)\coth\!\left(\frac{\beta\Omega_R}{2}\right).
\end{align}
In the thermodynamic limit, this approaches $k_{B\to D}\to J(\Omega_R)\coth(\beta\Omega_R/2
)$.

\section{Dynamics in the Markovian Limit}

In the limit of $\gamma \rightarrow \infty$, instead of employing the full HEOM formalism in Eq.~(9) of the main text, we adopt the truncated HEOM retaining only the Ishizaki–Tanimura truncation term~\cite{ishizaki05}. In this approximation, the dynamics reduces to
\begin{align}
    \frac{d}{dt}\rho 
    = -i\mathcal L \rho 
    - \sum_i \Delta_i \left[\sigma_i^z,\left[\sigma_i^z, \rho\right]\right],
\end{align}
where $\Delta_i = \sum_k \frac{d_k}{\gamma_k}$, with $d_k$ and $\gamma_k$ being the decomposition parameters of the bath correlation function defined in Eq.~(10) of the main text. 
This equation corresponds to the Markovian limit of the Redfield master equation.

\bibliography{quantum}